\documentstyle[psfig]{mn}

\newif\ifAMStwofonts

\def\lapp{\ifmmode\stackrel{<}{_{\sim}}\else$\stackrel{<}{_{\sim}}$\fi}
\def\gapp{\ifmmode\stackrel{>}{_{\sim}}\else$\stackrel{>}{_{\sim}}$\fi}
\def\psr{PSR~B1641$-$45}

\title{Single pulses from PSR B1641--45}
\author[Simon Johnston]
{Simon~Johnston\\
School of Physics, University of Sydney, NSW 2006, Australia.
}
\date{\today}
\pagerange{\pageref{firstpage}--\pageref{lastpage}}
\pubyear{2002}
\begin{document}
\maketitle
\label{firstpage}

\begin{abstract}
The integrated profile of \psr{} can be decomposed into
four Gaussian components. The sum of these Gaussian components can
be made to replicate
the total intensity, linear and circular polarization of
the integrated profile, and the discontinuity in the swing of the 
position angle.  Surprisingly, the single pulses from 
\psr{} can also be decomposed into these same Gaussian components
with only the amplitude of the components as a free parameter.
The distributions of the fitted amplitudes are log-normal.
Under the assumption that emission consists of 100\% polarized orthogonal
modes which are emitted simultaneously,
I show that the polarization properties of
the single pulses can also be replicated in a simple way. This lends
support to models involving the superposition of orthogonal modes rather than
disjoint emission of the modes.
\end{abstract}


\section{Introduction}
The integrated pulse profile from pulsars is one of the key
characteristics of the emission process.
Although single pulses vary widely both in intensity and shape, integrating
a few thousand pulses produces a highly stable integrated profile.
The integrated profile therefore yields clues as to the permanent
structures associated with the pulsar magnetosphere.
Kramer et al. (1994)\nocite{kwj+94} showed that the integrated profile
can be decomposed into a small number of Gaussian components.
These components represent a 2-D Gaussian
region of emission from a fixed height in the patchy pulsar beam \cite{lm88}.
In single pulses, some fraction of the component
is emitting when the beam sweeps the line of sight
and therefore the single pulses tend to have
narrower features (sub-pulses) than those of the integrated profile.
The sub-pulse features may also rotate about the magnetic axis,
producing the drifting sub-pulse phenomenology \cite{dr99}.
Integrated profiles show high levels of linear polarization and 
moderate circular polarization. Single pulses tend to be more 
highly polarized
than the integrated profile. The smooth position angle swing in
integrated profiles as a function of pulse longitude 
can occasionally be broken by 90\degr\ jumps which shows
that the emission is present in two orthogonal modes of radiation.
There has been considerable debate in the literature over whether
these two orthogonal modes are emitted simultaneously or disjointly.
Recently, McKinnon \& Stinebring (1998)\nocite{ms98} and
Karastergiou et al. (2002)\nocite{kkjl+02} have shown 
that simultaneously occuring modes nicely explain many 
aspects of single pulse data.

\psr{} is one of the brightest pulsars in the sky at 1.4 GHz.
It has a pulse period of 455~ms and
its pulse profile consists of a single component preceeded
by a low amplitude `precursor'.
Previous observations of this pulsar at a frequency of 1.6 GHz by Manchester,
Hamilton \& McCulloch (1980)\nocite{mhm80} showed that it had
moderate linear and circular polarization and that the leading `precursor'
had a high degree of linear polarization orthogonal to the 
polarization of the main pulse emission. Rankin (1990)\nocite{ran90} and
van Ommen et al. (1997)\nocite{vdhm97}
classify the pulsar as a core single even though it has a relatively
shallow position angle swing across the pulse.

\section{Observations and Data Reduction}
As part of a project investigating single pulses from a large sample
of pulsars, \psr{} was observed on 
2000 March 16 and 17 using the 64-m Parkes
radio telescope. The centre frequency, $\nu$, of the observations was
1413 MHz; at this frequency the system equivalent flux
density is 26 Jy.  The receiver consists of two
orthogonal feeds sensitive to linear polarization. The signals are
down-converted and amplified before being passed into the backend. The
backend, CPSR, is an enhanced version of the Caltech Baseband Recorder
(Jenet et al. 1997).  It consists of an analogue dual-channel
down-converter and digitizer which yields 2-bit quadrature
samples at 20 MHz.  The data stream is
written to DLT for subsequent off-line processing allowing all 4 Stokes
parameters to be computed. On the two occasions, 3102 and 3718
single pulses were recorded.
Before each observation, a 90-s observation of a pulsed
signal, directly injected into the receiver at a 45\degr\ angle to the feed,
is made. This enables the gain of each channel and the phase difference
between the channels as a function of frequency to be
calibrated. Observations of the flux calibrator Hydra~A were made
which allows absolute fluxes to be obtained.
\begin{figure}
  \centerline{\psfig{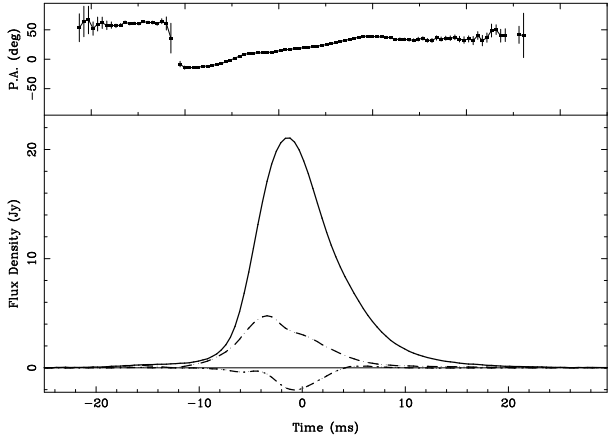}}
  \caption{Integrated pulse profile at 1.4~GHz. Position angle is
    shown on top, and the total intensity (solid line), linear (dashed line)
    and circular (dash-dot line) polarizations are shown in the bottom panel.
    The location of time zero is arbitrary, but roughly aligns with the
    pulse peak.}
  \label{profile}
\end{figure}
\begin{figure}
  \centerline{\psfig{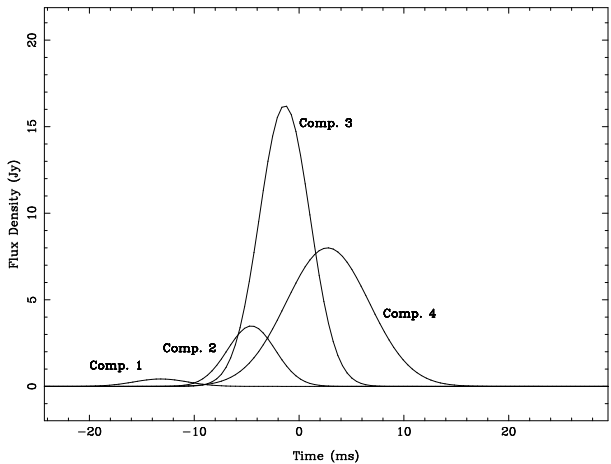}}
  \caption{The four Gaussians which make up the integrated profile
           of \psr{}. The parameters are given in Table~1.}
  \label{fits}
\end{figure}

The data were processed off-line using a workstation cluster at the
Swinburne Supercomputer Centre. Data reduction involves
coherent de-dispersion (Hankins \& Rickett 1975)
and includes quantization error corrections as described by
Jenet \& Anderson (1998). The data are folded at the apparent
topocentric period of the pulsar and the full Stokes profiles for each
pulse are written to disk. Flux calibration and instrumental calibration
are then carried out using information contained in the observation
of the pulsed (calibration) signal. The data in each frequency channel
is corrected for the rotation measure of the pulsar and all the
frequency channels are then summed to produce the final profile.
The final profiles consist of 1024 time-bins per pulse period 
for an effective time resolution of 0.44 ms.

\section{Integrated Profile}
\begin{figure}
  \centerline{\psfig{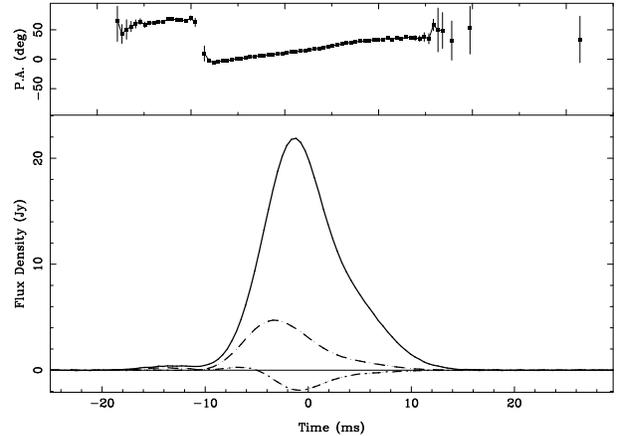}}
  \caption{The simulated profile of \psr{} constructed from the sum of
the four Gaussians parameterised in Table~1.}
  \label{simul}
\end{figure}
The integrated profile of \psr{} is shown in Figure~\ref{profile}.
The continuum flux density is 430~mJy and the peak of the profile has a flux
density of 21 Jy.
The pulsar shows a moderate amount of linear polarization ($\sim$20\%)
and a small amount of circular polarization ($\sim$6\%), which
is predominantly negative throughout.
The position angle swings by about 50\degr\ throughout the main pulse.
There is a `pedestal' or `pre-cursor' at the leading edge 
of the profile and the polarization
of this feature is orthogonal to the rest of the pulse.
The peak of the linear polarization occurs earlier than the peak in
total intensity. These results are consistent with the observations of
Manchester et al. (1980)\nocite{mhm80}.
There is scatter broadening of the pulse profile at this observing frequency
although it is difficult to measure without knowing the underlying
pulsar profile.
As part of a separate study of this pulsar (Rickett et al. In preparation),
we measured a pulse broadening time of 56~ms at an observing frequency of
660~MHz. This is an improvement over the only value currently in the
literature, 40$\pm$10~ms at 750~MHz by Komesaroff et al. (1973)\nocite{kac+73}.
Extrapolating to $\nu$=1413~MHz by assuming
that the broadening time scales as $\nu^{-4.4}$ yields a scattering time
of $\sim$2.0~ms at the observing frequency used here. This time is small
compared to the duration of the pulse, but likely manifests itself as
the tail of the emission at times beyond 13~ms in Figure~\ref{profile}.

After accounting for the orthogonal mode jump, the rotating vector
model (RVM) can be fitted to the position angle swing. As is commonly
the case, the angles  $\alpha$ (between the rotation and magnetic axes)
and $\beta$ (the offset of the line of sight from the magnetic axis)
are not well constrained, mostly because emission only occurs from 
a small fraction of pulse longitude. I adopt the parameters
$\alpha=30\degr$, $\beta=8\degr$ with the magnetic pole crossing located at
phase 0.931, just prior to the pulse peak. These parameters
provide an adequate fit to the data, and their exact values are not
crucial to the discussion below.

\begin{table}
\begin{tabular}{cccccc}
\multicolumn{1}{c}{No} & \multicolumn{1}{c}{Amplitude} &
\multicolumn{1}{c}{Location} & \multicolumn{1}{c}{Sigma} &
\multicolumn{1}{c}{Linear} & \multicolumn{1}{c}{Circular}\\
& \multicolumn{1}{c}{(Jy)} & \multicolumn{1}{c}{(ms)} & \multicolumn{1}{c}{(ms)}
& \multicolumn{1}{c}{fraction} & \multicolumn{1}{c}{fraction} \\
\hline
1 & 0.43  & --13.2 & 2.55  & --0.6 & 0.0 \\
2 & 3.49  & --4.55 & 2.28  & 0.9  & 0.2 \\
3 & 16.22 & --1.36 & 2.41  & 0.15 & --0.12 \\
4 & 8.00  &  +2.73 & 3.96  & 0.13 & --0.05 \\
\hline
\end{tabular}
\caption{Parameters of the components of \psr{} after Gaussian decomposition.
The location of the peaks is relative to zero time as shown in Figure~1.}
\end{table}
Kramer et al. (1994)\nocite{kwj+94} demonstrated the method of
decomposing the total intensity profiles of
pulsars into Gaussian components and showed that typically fewer than
five were necessary in the majority of cases.
Applying this method to \psr{}, I find that four Gaussian components
provide the best fit to the data. The fit
is excellent apart from the the trailing edge of the pulsar where
scatter broadening of the profile is clearly seen.
I define a Gaussian by
\begin{equation}
y = A\,\,\, e^{-\frac{1}{2}(\frac{x-\bar{x}}{\sigma})^2}
\end{equation}
and list the parameters of the fitted Gaussians in Table~1.
The fitted components are shown in Figure~\ref{fits}.  The
smallest amplitude component is necessary to fit the precursor. The
location of the second Gaussian is close to the peak of the linear
polarization. I therefore attempted to fit both the linear 
and the circular polarization
by varying the fractional polarization in each of the four
components. The results are listed in Table~1.
Note that the linear polarization of the precursor
is negative which implies that it is orthogonally polarized
to the rest of the emission.
The position angle at a given longitude is computed from the parameters 
of the RVM fit given above. The appropriate values of Stokes $Q$ and $U$
are then computed from the position angle for each Gaussian. These
are then summed to produce the integrated Stokes $Q$ and $U$ which are
then re-converted back to position angle.
The profile of the `simulated' pulsar is shown in Fig.~\ref{simul}.
The simulated profile  
can be compared directly with Fig.~\ref{profile}; in particular
the simulated profile nicely reproduces most of the main features of
the real profile, including the location of the orthogonal mode jump
which occurs when the amplitude of Gaussian 1 drops below that of
Gaussian 2. Also, both the linear and circular polarizations match well
with the true values.

\section{Single Pulses}
\subsection{Total Intensity}
\begin{figure}
  \centerline{\psfig{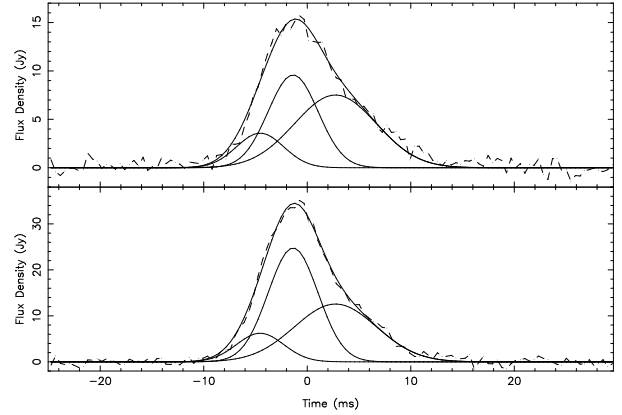}}
  \caption{Examples of single pulse fitting. Both panels show the observed
data with a dashed line. The three fitted Gaussians are shown with a solid
line as is the sum of these Gaussians. Top panel shows a strong pulse
whose components have amplitudes 6.2, 24.7 and 12.6 Jy. Lower panel shows a
weak pulse with component amplitudes 3.6, 9.6 and 7.5 Jy.}
  \label{singles}
\end{figure}
\begin{figure}
  \centerline{\psfig{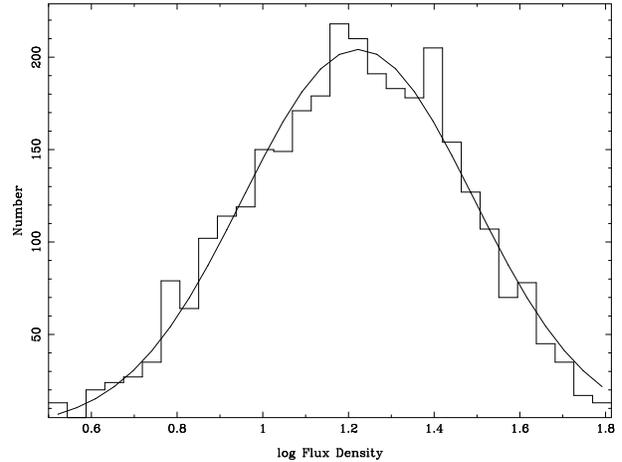}}
  \caption{Flux distribution of Gaussian~4 fitted to the
           single pulse data. Best fit is shown as a solid line.}
  \label{amps}
\end{figure}
The single pulses from this pulsar are rather featureless. They
generally show a similar morphology to that of the integrated profile and
there is no evidence for microstructure or drifting sub-pulses
as is the case for many other pulsars. The precursor component seen
in the integrated profile is rarely seen in the single pulses
as the noise dominates its emission.

Given the similarity between the single pulses and the integrated
profile, I attempted to perform Gaussian decomposition on the single pulses.
Most recently, Gupta \& Gangadhara (2003)\nocite{gg03} have 
applied a `window-thresholding'
technique to single pulse data. This allowed them to detect extra
components in the pulse profile which they then Gaussian fitted to obtain
their parameters. Here, I adopt a slightly different technique.
The width and centroid of the gaussians are fixed to
the values obtained from the fit to the integrated profile 
as it seemed apparent from visual inspection
that the single pulses had the same pulse width as the integrated profile.
However, the low amplitude Gaussian necessary to fit the precursor region
of the integrated profile was not included as it is too weak to
be seen in single pulses and thus could not be constrained in the 
fitting process.
I therefore performed fitting to the single
pulses with three free parameters (the amplitude of the
three Gaussians listed as 2-4 in Table~1).
This provided good fits (as judged by the $\chi^{2}$ statistic)
to the single pulse
data in more than 95\% of the cases. The majority of the failures came
when the flux density of the pulsar was low.
Often, the amplitude of Gaussian 2 was very low and the single pulses
were just as adequately fit with two Gaussians. However, some pulses
required a high amplitude for Gaussian 2.

The scatter broadening of
the pulse is not taken into account in the fitting procedure in
that symmetrical Gaussians rather than scatter broadened Gaussians are used.
The scattering only affects the tail of the profile, where the signal
to noise ratio is low. However, the goodness of fit is weighted towards the high
signal to noise points (i.e. the pulse peak). The fitting procedure is
therefore largely unaffected by the scattering. I return to the effects
of scattering on the implications of the results in a later section.
Figure~\ref{singles} shows two examples of single pulses and the 
fitted Gaussians. The fits are excellent and show that the scatter
broadening is not playing a significant role in 
determining the fitted parameters.

The question then arises as to the distribution of the amplitudes
of the Gaussians. Figure~\ref{amps} shows the distribution for
Gaussian~4. The distribution is clearly log-normal with a mean
amplitude of 16.7 Jy and a $\sigma$ of 0.27 (in the log).
Similarly, Gaussian~3 is log-normally distributed with a mean
of 38.7 Jy and a $\sigma$ of 0.13 (in the log).
The statistics of Gaussian~2 are more problematical to determine
as often this component is weak and hence dominated by noise.
A reasonable fit is obtained with a log-normal distribution with
a mean of 7.45 Jy and a $\sigma$ of 0.3 in the log.
Note that the means obtained in the fitting process correspond nicely
to the peak amplitudes obtained during the Gaussian decomposition
process on the integrated profile. This gives confidence that the
fitting procedure is working correctly.
Remarkably, therefore, the single pulses from this pulsar can be
reproduced by drawing three random numbers from the above distributions
and summing the resultant Gaussians together.
\begin{table*}
\begin{tabular}{ccccccc}
\multicolumn{1}{c}{No} & \multicolumn{1}{c}{log(Amp 1)} &
\multicolumn{1}{c}{log(Amp 2)} & \multicolumn{1}{c}{Sigma} &
\multicolumn{1}{c}{Corr Coeff} &
\multicolumn{1}{c}{Circular fraction} & \multicolumn{1}{c}{Circular spread}\\
\hline
1 & --0.189 & 0.421 & 0.00  & 0.7 & 0.0 & 0.15 \\
2 & 0.451 & --0.829 & 0.27 & 0.7 & 0.2 & 0.15 \\
3 & 0.941 & 0.811 & 0.13  & 0.7 &  --0.12 & 0.15 \\
4 & 0.571 & 0.461 & 0.27  & 0.7 & --0.05 & 0.15 \\
\hline
\end{tabular}
\caption{Parameters used in the simulation of single pulses from \psr{}}
\end{table*}
\begin{figure}
  \centerline{\psfig{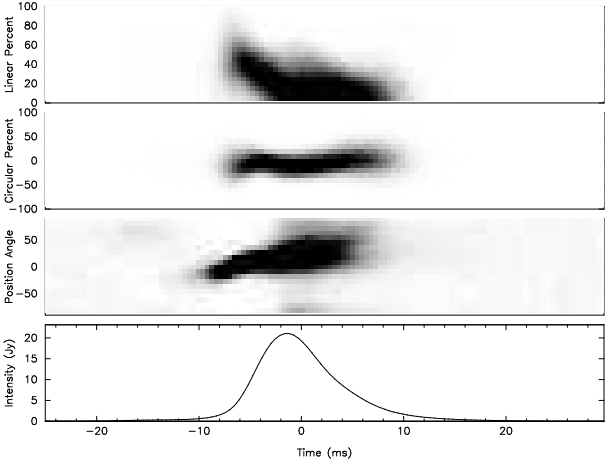}}
  \caption{Grey-scale representation of the polarization properties of
the single pulses from \psr{}. Black denotes regions of high counts.}
  \label{grey}
\end{figure}
\subsection{Polarization}
In order to examine the polarization of the single pulses,
an analysis similar to that of
Stinebring et al. (1984a,b)\nocite{scr+84,scw+84} was performed.
For the linear polarization, for example, a two dimensional grid 
is constructed with the pulse phase along the x-axis and 20 bins along the
y-axis from 0 to 100\%. For each pulse the appropriate cell is then
updated by one depending on its linear polarization. The same procedure 
is carried out for the circular polarization and the position angle.
The values in the cells are then converted to a linear
grey-scale between white
and black depending on the counts in that cell.
Figure~\ref{grey} shows the grey-scale representation of the data.
The figure shows that although there is a large spread in 
the linear and circular polarization, rather few pulses have
a net polarization greater than 50\% apart from the leading edge
of the profile. There is also little evidence of orthogonally 
polarised modes in the position angle distribution.

Figure \ref{b960} shows the polarization properties of the phase bin where 
the linear polarization is maximum. In this bin, one mode dominates
virtually all the time, the position angle distribution is narrow around its
mean value. The fractional linear polarization is high and the
circular polarization shows large flux densities of either sign although
there is a slight preference for negative values.
This contributes to give a low net circular polarization.

Figure~\ref{b976} shows the same information for a phase bin on
the trailing edge of the profile. Both the total intensity and linearly
polarized intensity are lower as is the fractional linear polarization.
The PA distribution is much broader than seen in Fig.~\ref{b960}.
This is partly due to increased uncertainity in the measurement of the PA
(as the linear polarization is lower) and partly due to the presence
of the orthogonal mode appearing near PA $-80\degr$.
The circular polarization is more evenly distributed around zero.

\section{Simulating Single Pulses}
I have already demonstrated that replicating the single pulses in
total intensity is a relatively simple task for this pulsar. It
involves drawing three random numbers from a log-normal distribution.
What further parameters need
to be added in order to replicate the polarization of the single pulses?

As a starting point I use the model proposed by
McKinnon \& Stinebring (1998)\nocite{ms98}. In their model both
polarization modes are emitted simultaneously and are 100\% polarized.
Then, the total power is the sum of the intensities of the two modes
and the linear polarization is their difference.
The amplitudes of the modes are highly correlated.
They found that, under these assumptions, the single pulse data of
Stinebring et al. (1994) could be replicated.
McKinnon \& Stinebring (1998)\nocite{ms98} treated each phase
bin in an independent fashion.
In a further paper, they also showed that the integrated profile
could be `split' into the two orthogonal modes \cite{ms00}.
This involved a simple mathematical manipulation based on the
fractional linear polarization in each bin (see also Karastergiou
et al. 2003).

In this instance, I assume that the Gaussians described in Section~3
are a coherent emitting unit, or patch, and that they emit radiation
simultaneously in two, 100\% polarized orthogonal modes.
Ignoring for now the circular polarization, one can apply
the technique of McKinnon \& Stinebring (2000) to
compute the mean amplitude of each of the modes for each of the Gaussians
under these assumptions using the data from Table 1 and knowing
the integrated fractional polarization. Furthermore, the
distribution of the amplitudes of these Gaussians are also known
from the fitting described above; they are all log-normal.
\begin{figure*}
  \centerline{\psfig{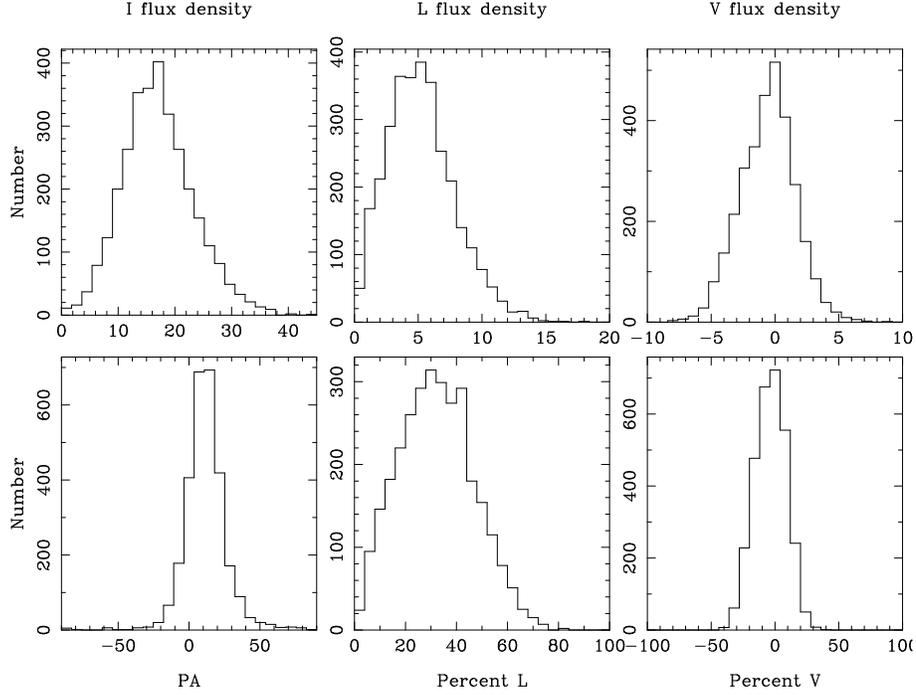}}
  \caption{Polarization properties of bin at time --3.4~ms.
Flux density in Jy, position angle in degrees. The 1-$\sigma$ noise
is level is 0.6 Jy.}
  \label{b960}
\end{figure*}
\begin{figure*}
  \centerline{\psfig{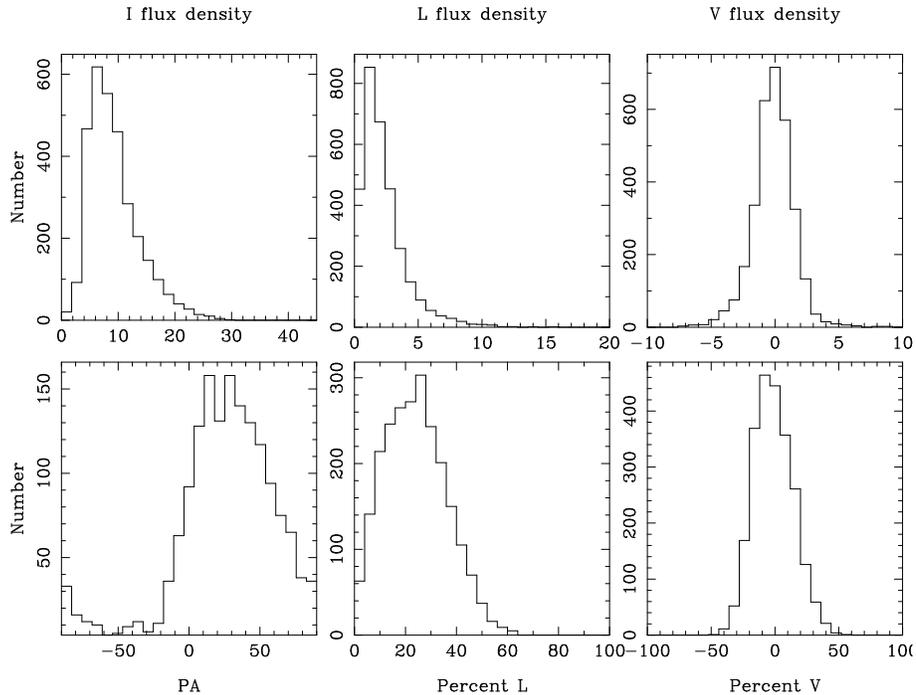}}
  \caption{Polarization properties of bin at time 3.64~ms. Flux density in Jy,
position angle in degrees. The 1-$\sigma$ noise is level is 0.6 Jy.}
  \label{b976}
\end{figure*}

In order to simulate the polarization of the single pulses from \psr{},
therefore, we now need to draw 8 random numbers (one pair for each 
of the Gaussians in the table) 
from a log-normal distribution with the mean 
and sigma as given in Table~2.
Note that the random values for each mode are correlated
according to the correlation coefficient given in the table
\cite{ms98}. Also, the parameters for Gaussian 1 are not well
constrained, but do not play a major role in the single pulse data as
the flux density is low.
The total intensity of each Gaussian is then the 
sum of the amplitudes of the two orthogonal modes and the linear
polarization is the difference of their amplitudes.
This accounts for the linear polarization but does not include any
circularly polarized component.
The mean circular polarization of each individual Gaussian is already known
from the fits to the integrated profile. It is tempting to associate
a handedness of circular polarization with a given orthogonal mode
as has been proposed by e.g. Cordes, Rankin \& Backer (1978)\nocite{crb78}.
However, this seems be to true only at observing frequencies lower than
$\sim$600~MHz. Recent observations of the Vela pulsar \cite{kjv02} and
PSR~B1133+16 \cite{kjk03} show that the situation
is much more complex at frequencies above about 1~GHz. There is a lack
of correlation between the handedness of circular polarization and
the dominant orthogonal mode, but a rather large spread in the values
of both hands of circular polarization.
For each component (not each orthogonal mode), I therefore draw the 
circular polarization from a Gaussian distribution with a mean and
standard deviation as listed in Table~2.
\begin{figure*}
  \centerline{\psfig{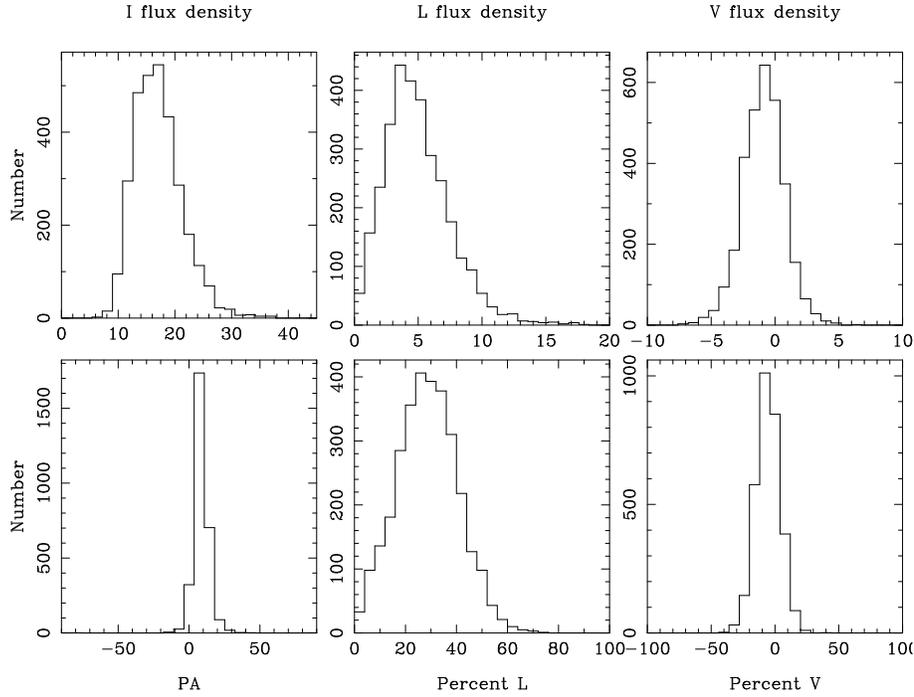}}
  \caption{Simulated data from the same time bin as in Fig~\ref{b960}.}
  \label{s960}
\end{figure*}
\begin{figure*}
  \centerline{\psfig{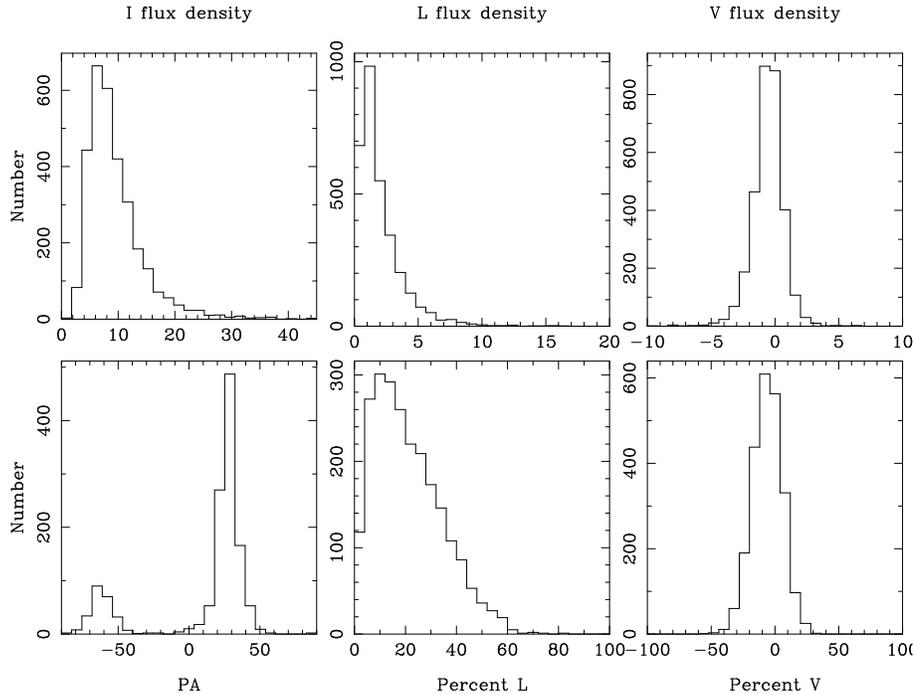}}
  \caption{Simulated data from the same time bin as in Fig~\ref{b976}.}
  \label{s976}
\end{figure*}

Single pulses simulated in such a way bear a strong resemblance to the
true single pulse data. Figs~\ref{s960} and \ref{s976} show the simulated
data for the same phase bins as Figs~\ref{b960} and \ref{b976}.
Comparison between the total intensity distributions show that the 
low flux density pulses are under-represented, however the high intensity
data are well matched. The linear and circular polarizations are 
well replicated. The biggest discrepancy lies in the spread of
the position angle distributions. In the simulated data, the spread is
relatively narrow and the orthogonal modes are well separated. The
spread in values is almost entirely due to receiver noise as I have assumed that
the position angle is fixed for a given bin. In the real data the spread
is much larger than a delta function convolved with the noise.
This is similar to the results found by
McKinnon \& Stinebring (1998)\nocite{ms98} and
Karastergiou et al. (2002)\nocite{kkjl+02}.

\section{Discussion}
I have shown that Gaussian decomposition of the integrated
profile not only provides a good fit to the total intensity but
can also fit the polarization profile, including the location
of the orthogonal jump. In this pulsar, at least, these same Gaussians
can be made to fit the single pulse data with only their amplitudes
as a free parameter. Interestingly the distribution of these
amplitudes is log-normal. Log-normal distributions are seen both
in the time-averaged flux density of pulsars
and in phase resolved statistics \cite{cjd01,cjd03}. The implications
are then that pulsar emission may be interpreted in terms of
stochastic growth theory and that the emission mechanism is linear,
involving either a direct linear instability or linear mode
conversion of non-escaping waves driven by a linear instability as
discussed in detail in Cairns et al. (2003)\nocite{cjd03}.

Under the assumptions of 100\% polarized
orthogonal modes, the polarization distributions of the 
single pulse data can also be reproduced
from a relatively small number of random variables.
This extends the work of McKinnon \& Stinebring (2000)\nocite{ms00}
by implying that the sub-pulses can be characterised in such a
way (not just a particular phase bin) and that log-normal statistics
are clearly the dominant statistic for the observed flux density
distributions. Any physical model must therefore have as its basis
an emission beam which is 100\% polarized and which is 
subsequently split through e.g.  bi-refringence into two orthogonally
polarized modes.
However, it is clear that propagation effects, not taken into account in the
simulation process, strongly affect the position angle and
the degree of circular polarization seen in single pulses.
This situation has also been seen previously by a number of
observers and discussed in the context of theoretical models
by Petrova (2001)\nocite{pet01}. In her model, the polarization
modes recombine in the `polarization limiting region'. This mode
coupling then allows for significant deviations in the position
angle and in the circular polarization, with greater position
angle deviations leading to increased circular polarization.
The data here do not entirely support this picture. Comparison between
Figs \ref{b960} and \ref{b976} show that the former has a much narrower
spread of position angles than the latter but that the percentage
V distributions look very similar. The effects of refraction in the
magnetosphere are also likely to be important as the two modes are
affected differently. This is likely to result in a mode mixture
at a given observer's longitude and such an effect would also serve to
broaden the position angle distribution.

The applicability of this method to other pulsars is not clear.
For example, in the Vela pulsar, Krishnamohan \& Downs (1983)\nocite{kd83}
showed that four Gaussian components fitted the total intensity data
and could account for some of the single pulse data. More
recent observations with much higher time resolution \cite{kjv02}
showed that microstructure was prevalent in the single pulse data.
Clearly, no amount of Gaussian fitting will reproduce microstructure.
Other pulsars, such as PSR B0950+08 show a bewildering variation
in the amplitude and location of single pulse components, and there
is almost no resemblance between the single pulses and the integrated
profile. Many pulsars, such as PSR B1133+16, show orthogonal mode
jumps near the peak of a pulse component and again such a feature
is hard to reproduce using simple Gaussian decomposition.

However, it may be that there is a class of pulsars similar to \psr{}
for which this method will work very well. One possibility is that
the scatter broadening in this pulsar blurs the features with
time-scales less than the sub-pulse time. The same was also true
of the low resolution observations of the Vela pulsar where the
dispersion measure smearing also blurred out the microstructure.
Processing techniques which involve smoothing over the microstructure
are also possible. For example, van Leeuwen et al. (2002)\nocite{vlk+02}
fitted Gaussian components
to PSR B0809+74 in order to study the drift bands and noted how similar
the Gaussians were in amplitude and width.

It therefore appears to be possible to get a handle on processes
on the scale of tens of milliseconds (the subpulses) without delving
into the complexities which occur on timescales significantly shorter.
The data seem to favour the idea of simultaneously occuring
modes of emission, with propagation through the magnetosphere influencing
the degree of circular polarization and the position angle variations.

\section*{Acknowledgments}
The Australia Telescope is funded by the Commonwealth of 
Australia for operation as a National Facility managed by the CSIRO.
I thank W.~van Straten for help with the data reduction,
A.~Karastergiou for useful discussions and the referee for improving
the content of the paper.

\bibliography{modrefs,psrrefs,crossrefs}
\bibliographystyle{mn}
\label{lastpage}
\end{document}